\documentclass[aps,twocolumn,superscriptaddress,prl]{revtex4-2}

\usepackage{bm,graphicx,textcomp,amssymb,amsmath,dcolumn,color}

\usepackage{subcaption}

\newcommand{\ket}[1]{\left|#1\right>}
\newcommand{\bra}[1]{\left<#1\right|}

\newcommand{\nn}{\nonumber\\}

\newcommand{\f}[1]{\mbox{\boldmath$#1$}}

\newcommand{\bea}{\begin{eqnarray}}
\newcommand{\ea}{\end{eqnarray}}
\newcommand{\eea}{\end{eqnarray}}

\begin{document}

\title{Higher harmonics in Mott-Hubbard insulators as sensors} 



\author{Abdelrahman~Azab}

\affiliation{Helmholtz-Zentrum Dresden-Rossendorf,
Bautzner Landstra{\ss}e 400, 01328 Dresden, Germany,}

\author{Friedemann~Queisser}

\affiliation{Helmholtz-Zentrum Dresden-Rossendorf,
Bautzner Landstra{\ss}e 400, 01328 Dresden, Germany,}

\author{Gulloo Lal Prajapati}

\affiliation{Helmholtz-Zentrum Dresden-Rossendorf,
Bautzner Landstra{\ss}e 400, 01328 Dresden, Germany,}

\author{Jan-Christoph Deinert}

\affiliation{Helmholtz-Zentrum Dresden-Rossendorf,
Bautzner Landstra{\ss}e 400, 01328 Dresden, Germany,}

\author{Ralf Sch\"utzhold}

\affiliation{Helmholtz-Zentrum Dresden-Rossendorf,
Bautzner Landstra{\ss}e 400, 01328 Dresden, Germany,}

\affiliation{Institut f\"ur Theoretische Physik,
Technische Universit\"at Dresden, 01062 Dresden, Germany,}

\date{\today}

\begin{abstract}
Using strong-coupling time-dependent perturbation theory, we study the response of Mott and charge-transfer insulators to an oscillating electric field. We derive analytical expressions for the resulting higher-harmonic currents and show that they encode information about spin order and microscopic hopping pathways. The results demonstrate that higher harmonics can serve as probes of correlated materials and as sensors of the applied driving field.
\end{abstract}

\maketitle 
\section{Introduction}
Higher harmonics constitute a powerful probing phenomenon. They enable quantum-nondemolition (QND) measurements \cite{hhgqnd,hhgqnd2} and have been employed as sensors across diverse regimes, including biological systems \cite{hhgbio}, organic molecules \cite{hhgdipole}, and weakly interacting systems and materials \cite{hhgweakly1,hhgmaterial}.

In strongly interacting systems, however, the same sensing concepts remain comparatively unexplored due to their intrinsic many-body complexity. Motivated by recent experimental studies of higher harmonics in correlated materials \cite{anomalous, experiment}, theoretical efforts have begun to explore the underlying mechanisms, with the Fermi–Hubbard model providing a canonical framework for such systems\cite{hhgfriedemann}. While mean-field approaches have been applied to higher-harmonic generation in Mott insulators \cite{hhgmft, dmft}, a transparent analytical understanding of the microscopic origin of these harmonics and their relation to correlation effects is still underdeveloped.

In this work, we analyze driven Fermi–Hubbard systems and related models in the strong-coupling regime using time-dependent perturbation theory. We derive analytical expressions for higher-harmonic currents and show that higher-harmonics in Mott systems are spin dependent while in charge-transfer insulators, spin dependence is highly influenced by hopping channels and band filling. The resulting spectra further exhibit resonances at multiples of the interaction energy. This establishes higher harmonics as probes of correlated materials and as sensors in the strongly interacting regime.

The paper is organized as follows. We first present the higher-harmonic currents for single- and two-band Fermi–Hubbard models, followed by an analysis of charge-transfer insulators. We then discuss how the harmonic response reflects spin order and microscopic hopping pathways, enabling sensing of both material properties and the applied drive.
\section{Strong-Coupling Time-Dependent Perturbation Theory}

The main model considered in this work is the Fermi-Hubbard model ($\hbar=1$) \cite{fermihubbard,hubbard}
\begin{equation}
\label{Fermi-Hubbard}
\hat H
=
-\sum_{\mu\nu s} T_{\mu\nu}
\hat c_{\mu s}^\dagger\hat c_{\nu s}
+U\sum_\mu \hat n_\mu^\uparrow\hat n_\mu^\downarrow \,,
\end{equation}
where $\hat c_{\mu s}^\dagger$ and $\hat c_{\mu s}$ are fermionic creation and annihilation operators for spin 
$s\in\{\uparrow,\downarrow\}$ at lattice site $\mu$, and 
$\hat n_\mu^s=\hat c_{\mu s}^\dagger\hat c_{\mu s}$ are the corresponding number operators. 
The hopping matrix $T_{\mu\nu}$ is taken to be $T$ for nearest neighbors and zero otherwise, while $U$ denotes the on-site Coulomb repulsion. 

In the strong-interaction regime ($U\gg T$) at half filling, the ground state is a Mott insulator with one electron localized on each lattice site. This is due to the suppression of charge fluctuations by the small parameter $T/U$ \cite{attractionvsrepulsion, mott}. 

To perform an expansion in $T/U\ll 1$, we separate the Hamiltonian into $\hat{H}=\hat{H}_0+\hat{H}_T$. Where $\hat{H}_0$ is the unperturbed Hamiltonian with ground state $\ket{\psi_0}$ and $\hat{H}_T$ is taken as perturbation which is usually of order $T$. This allows us to expand the state of the system to second order as
\begin{align}
\label{Perturbation}
\ket{\psi}
=\ket{\psi_0}-i\int_0^t dt'\hat{H}_I(t')\ket{\psi_0}\notag\\-\int_{0}^t dt_1\int_0^{t_1}dt_2\hat{H}_I(t_1)\hat{H}_I(t_2)\ket{\psi_0}
\,,
\end{align}
where $\hat{H}_I(t')=e^{i\hat{H}_0t'}\hat{H}_T(t')e^{-i\hat{H}_0t'}$ is the time dependent perturbation Hamiltonian in the interaction picture.

Expectation values of observables are evaluated using time-dependent operators. In particular, the current operator in the interaction picture is
\begin{equation}
\hat J_{\mu\nu}(t)
=
i\sum_s\!\left(
\hat c_{\mu s}^\dagger(t)\hat c_{\nu s}(t)
-
\hat c_{\nu s}^\dagger(t)\hat c_{\mu s}(t)
\right),
\end{equation}
where the time-dependent Fermionic operators are
\[
\hat c_{\mu s}(t)=e^{i\hat H_0 t}\hat c_{\mu s}e^{-i\hat H_0 t},
\qquad
\hat c_{\mu s}^\dagger(t)=e^{i\hat H_0 t}\hat c_{\mu s}^\dagger e^{-i\hat H_0 t}.
\]
With a given state and the interaction-picture operators defined above, the time-dependent current can be calculated.
\section{Single-band Driven Fermi-Hubbard Model}
To study the generation of higher harmonics, we first consider the driven single-band Fermi-Hubbard model, described by
\begin{align}
\hat H
=
-\sum_{\mu\nu s} T_{\mu\nu}
\hat c_{\mu s}^\dagger\hat c_{\nu s}
+U\sum_\mu \hat n_\mu^\uparrow\hat n_\mu^\downarrow
+\sum_{\mu s} V_\mu(t)\hat n_{\mu s},
\end{align}
where $V_\mu(t)=q\,\mathbf x_\mu\!\cdot\!\mathbf E(t)$ is the site-dependent scalar potential of an electron with charge $q$ in a spatially uniform electric field $\mathbf E(t)=\mathbf E\cos(\omega t)$.

Equivalently, the Hamiltonian can be written in the Peierls representation as \cite{hhgfriedemann}
\begin{align}
\hat H
=
-\sum_{\mu\nu s} T_{\mu\nu}(t)
\hat c_{\mu s}^\dagger\hat c_{\nu s}
+U\sum_\mu \hat n_\mu^\uparrow\hat n_\mu^\downarrow,
\end{align}
where the time-dependent hopping is
\[
T_{\mu\nu}(t)=T_{\mu\nu}e^{i\varphi_{\mu\nu}(t)}, 
\qquad 
\dot\varphi_{\mu\nu}(t)=q\,\mathbf E(t)\!\cdot\!\mathbf r_{\mu\nu},
\]
with $\mathbf r_{\mu\nu}=\mathbf x_\mu-\mathbf x_\nu$.

Separating the Hamiltonian into $\hat{H}_0=U\sum_\mu\hat n_\mu^\uparrow\hat n_\mu^\downarrow$ and $\hat{H}_T(t)=-\sum_{\mu\nu s} T_{\mu\nu}(t)
\hat c_{\mu s}^\dagger\hat c_{\nu s}$, the leading-order current in the Mott ground state is
\begin{align}
        \langle \hat{J}_{\mu\nu}(t)\rangle=i\int_0^t dt'\bra{\psi_0}(\hat{H}_I(t')\hat{J}_{\mu\nu}(t)-\hat{J}_{\mu\nu}(t)\hat{H}_I(t')\ket{\psi_0},
\end{align}
This gives the odd harmonic current in terms of Bessel functions $\mathcal{J}_m(\gamma)$
\begin{widetext}
\begin{align}\label{1bandhh}
     \langle \hat{J}_{\mu\nu}(t)\rangle\propto T_{\mu\nu}\left(
\frac14-\left\langle\hat{\f{S}}_\mu\cdot\hat{\f{S}}_\nu\right\rangle
\right)\sum_{\substack{m=0}}\mathcal{J}_{2m+1}(\gamma)\Bigg[\frac{U\sin((2m+1)\omega t)-(2m+1)\omega\sin(Ut)}{U^2-((2m+1)\omega)^2}\bigg],
\end{align}
\end{widetext}
where $\gamma=\frac{qEl}{\omega}$ with $l$ representing the lattice spacing. $\hat S^a_\mu=\sum_{ss'}\sigma_{ss'}^a
\hat c_{\mu s}^\dagger\hat c_{\mu s'}/2$ are the spin operators
at site $\mu$ with $\sigma_{ss'}^a$ being the usual Pauli matrices.

Focusing as an example on the third harmonic current, we obtain
\bea
\label{current}
J_{\mu\nu}^{3\omega}
\propto
{\cal N}\,
\frac{T_{\mu\nu}}{U}
\left(
\frac14-\left\langle\hat{\f{S}}_\mu\cdot\hat{\f{S}}_\nu\right\rangle
\right) 
\left(
\frac{q\f{E}\cdot\f{r}_{\mu\nu}}{\omega}
\right)^3  
\,.
\ea
This demonstrates a strong dependence of the higher-harmonic current on the underlying spin correlations. For ferromagnetic alignment, $\langle\hat{\f{S}}_\mu\cdot\hat{\f{S}}_\nu\rangle=1/4 $, the current vanishes due to Pauli blocking, which prevents doublon formation for parallel spins. In contrast, a paramagnetic configuration with $\langle\hat{\f{S}}_\mu\cdot\hat{\f{S}}_\nu\rangle=0 $ yields a finite current, which is further enhanced for antiferromagnetic correlations ($\langle\hat{\f{S}}_\mu\cdot\hat{\f{S}}_\nu\rangle<0$). For example, in an Ising antiferromagnet with $\langle\hat{\f{S}}_\mu\cdot\hat{\f{S}}_\nu\rangle=-1/4$, the current reaches a larger magnitude. Even stronger enhancement is possible in Heisenberg antiferromagnets, where spin correlations depend sensitively on lattice geometry \cite{heisenbergaf1,heisenbergaf2,heisenbergaf3}. This behavior, consistent with the experimental observations in \cite{experiment}, enables discrimination between different magnetic orderings as well as between lattice geometries in Heisenberg antiferromagnets.

\begin{figure}[t]
    \centering
    
    \begin{subfigure}{\columnwidth}
        \centering
        \includegraphics[width=\linewidth]{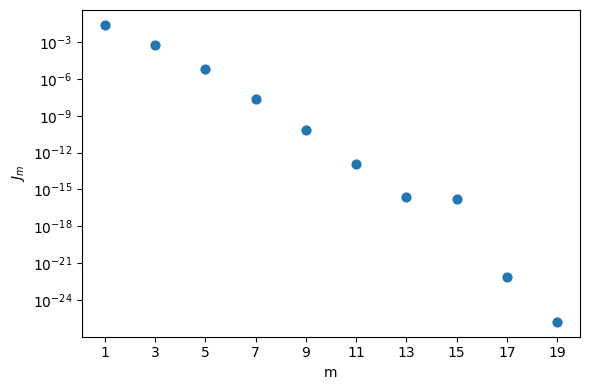}
        \label{fig:weak}
    \end{subfigure}
    
    \vspace{0.5em}
    
    \begin{subfigure}{\columnwidth}
        \centering
        \includegraphics[width=\linewidth]{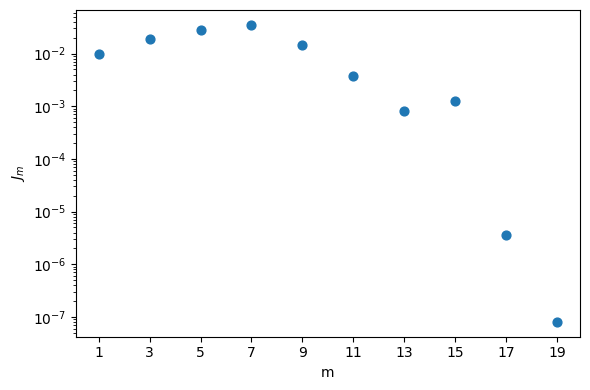}
        \label{fig:strong}
    \end{subfigure}
    
    \caption{Odd-harmonic spectrum showing the maximum current for each harmonic order $m$ for two electric-field strengths: weak field $Er_{\mu\nu}/U=0.05$ (top) and strong field $Er_{\mu\nu}/U=0.5$ (bottom). Parameters: $U=15\omega$, producing a resonance at harmonic $m=15$.}
    \label{fig:hhg_compare}
\end{figure}

Using Eq.~\eqref{1bandhh}, the harmonic spectrum can be computed for different driving strengths, as shown in Fig.~\ref{fig:hhg_compare}. The resulting spectra closely reproduce the behavior obtained within the hierarchy-of-correlations approach \cite{hhgfriedemann} and mean-field theory \cite{hhgmft,dmft}. The pronounced differences between the weak- and strong-field regimes allow immediate discrimination between them. In particular, the resonance condition $m\omega \approx U$ directly reveals the interaction energy, while the plateau position encode the field amplitude.

\section{Two-band Fermi-Hubbard Model}

We next consider a two-band extension of the Fermi-Hubbard model with two orbitals $A=1,2$ per lattice site $\mu$. The Hamiltonian is
\begin{equation}
\label{two-band-Fermi-Hubbard}
\hat H
=
-\sum_{\mu\nu AB s} T_{\mu\nu}^{AB}(t)
\hat c_{\mu A s}^\dagger\hat c_{\nu B s}
+
\sum_{\mu AB ss'}
U_{AB}^{ss'}
\hat n_{\mu A}^{s}\hat n_{\mu B}^{s'} ,
\end{equation}
where $T_{\mu\nu}^{AB}$ describes intra-band ($A=B$) and inter-band ($A\neq B$) hopping processes. Here, $U_{AB}^{ss'}$ represents the Coulomb repulsion energies, both in the same band, and in different bands. Where, the terms $U_{11}^{ss}$ and $U_{22}^{ss}$ correspond to the chemical potentials of the bands which are taken to have an energy gap $\Delta E=U_{22}^{ss}-U_{11}^{ss}>0$

\begin{figure}[t]
    \centering
    \includegraphics[width=1\linewidth]{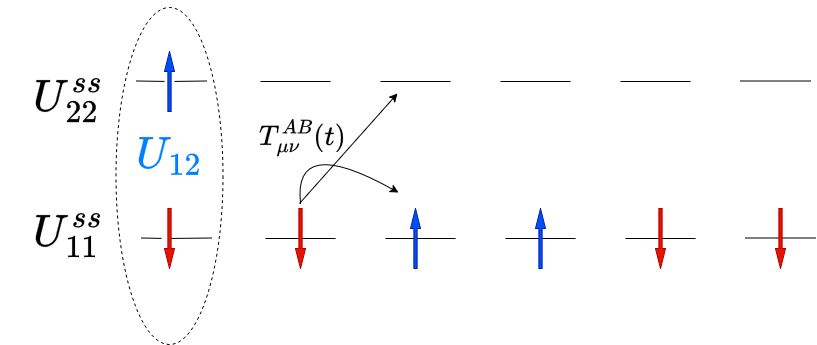}
    \caption{Two-band Fermi-Hubbard setup with lower Mott band at energy $U^{ss}_{11}$ and empty upper band at energy $U^{ss}_{22}$. The dotted ellipse shows the on-site Coulomb repulsion. Having two electrons in the same lattice site and the same band, would result in a Coulomb repulsion $U^{ss'}_{11}$ or $U^{ss'}_{22}$. Intra- and inter-band hoppings are denoted by $T_{\mu\nu}^{AB}(t)$.}
    \label{2orbit}
\end{figure}

We now take, as a ground state, the state with a half filled lower band (with one particle per site), and a completely empty upper band, as illustrated in Fig. \ref{2orbit}.

Proceeding analogously to the single-band case, we separate the Hamiltonian as $\hat H=\hat H_0+\hat H_T$, with
\[
\hat H_0=\sum_{\mu AB ss'} U_{AB}^{ss'} \hat n_{\mu A}^{s}\hat n_{\mu B}^{s'},
\qquad
\hat H_T=-\sum_{\mu\nu AB s} T_{\mu\nu}^{AB}(t)
\hat c_{\mu A s}^\dagger\hat c_{\nu B s}.
\]
Applying strong-coupling time-dependent perturbation theory yields higher harmonic currents with a structure analogous to Eq.~\eqref{1bandhh}. The third-harmonic contribution becomes
\begin{equation}
\label{current-double}
J_{\mu\nu}^{3\omega}
\propto 
\frac{T_{\mu\nu}^{11}}{U_{11}}
\left(
1-4\left\langle\hat{\mathbf S}_\mu^{1}\cdot\hat{\mathbf S}_\nu^{1}\right\rangle
\right) 
+
\frac{2T_{\mu\nu}^{12}}{U_{12}+\Delta E} ,
\end{equation}
where $\hat{\mathbf S}_\mu^{1}=\sum_{ss'}\boldsymbol{\sigma}_{ss'}
\hat c_{\mu 1 s}^\dagger\hat c_{\mu 1 s'}/2$ is the spin operator in the lower band. This expression assumes spin-independent Coulomb interactions. In the presence of spin-dependent interactions, the inter-band term would also acquire spin dependence. However, since the upper band is empty, only the spin correlations of the lower band contribute.

Thus, as in the single-band case, the harmonic response of intra-band processes is directly controlled by spin order, while inter-band hopping provides a spin-independent channel.
\section{Charge-transfer insulator}
A charge-transfer insulator is characterized by ligand $p$ states that lie energetically between the lower and upper Hubbard $d$ bands, with a charge-transfer gap $\Delta=\epsilon_d-\epsilon_p$ smaller than the on-site $d$ Coulomb repulsion $U$ \cite{CTdescription,zaanenallen,ctivsmh}. In this regime, charge fluctuations preferentially occur between $p$ and $d$ orbitals rather than within the $d$ band.

To describe this system, we consider the Hamiltonian
\bea
\label{charge-transfer}
\hat H
&=&
-\sum_{\mu\nu s} T_{\mu\nu}^d(t)
\hat d_{\mu s}^\dagger\hat d_{\nu s}
+U\sum_\mu
\hat d_{\mu\uparrow}^\dagger\hat d_{\mu\uparrow}
\hat d_{\mu\downarrow}^\dagger\hat d_{\mu\downarrow}
\nn
&&-\sum_{\mu\nu s} T_{\mu\nu}^p(t)
\hat p_{\mu s}^\dagger\hat p_{\nu s}
+\epsilon_p 
\sum_{\mu s}
\hat p_{\mu s}^\dagger\hat p_{\mu s}
\nn
&&-\sum_{\mu\nu s} T_{\mu\nu}^{pd}(t)
\left(\hat d_{\mu s}^\dagger\hat p_{\nu s}+
\hat p_{\mu s}^\dagger\hat d_{\nu s}\right) 
\,,
\ea
where the first line describes the Hubbard band, the second line 
the metallic $p$-band  and the third line 
is the overlap between both bands. 

\begin{figure}[t]
    \centering
    \includegraphics[width=1\linewidth]{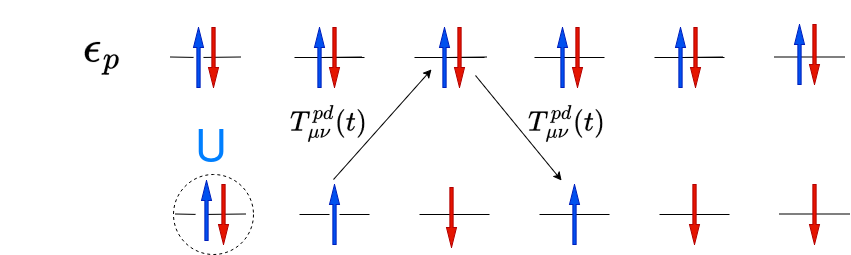}
    \caption{Ground state of charge-transfer insulator with a filled $p$ band and a Mott-localized $d$ band. The dotted circle represents the Coulomb repulsion when two electrons occupy the same lattice site in the $d$-band.}
    \label{CT}
\end{figure}

In this case, we treat the $T^{pd}_{\mu\nu}$ as a perturbation.
In the charge-transfer insulating limit, the $p$-band is taken to be completely filled while, as before, the $d$-band forms a Mott state, as illustrated in Fig. \ref{CT}.

In contrast to the Hubbard cases discussed above, the leading-order harmonic current arising from $p$–$d$ processes shows no spin dependence. This follows from the filled $p$ band: hopping between $p$ and $d$ orbitals requires an electron with spin opposite to that already occupying the $d$ site (Pauli blocking), but both spin species are always available in the full $p$ band. Consequently, the $p$–$d$ and $p$–$p$ hopping channels generate spin-independent harmonic currents. Spin dependence therefore appears only in the $d$–$d$ contribution, analogous to the single-band Hubbard case.

\section{Higher harmonics as sensors}
The harmonic response derived above provides two independent sensing channels: spin correlations of the material and the field-dependent structure of the harmonic spectrum.

For Hubbard systems, the harmonic amplitudes depend explicitly on nearest-neighbor spin correlations. Ferromagnetic order suppresses the signal, whereas antiferromagnetic correlations enhance it. Different antiferromagnetic states therefore produce distinct harmonic responses, enabling distinguishing between magnetic orderings and providing sensitivity to lattice geometry. This spin-dependent sensing mechanism applies to both single- and multi-band Hubbard models.

In contrast, for charge-transfer insulators the dominant $p$–$d$ and $p$–$p$ hopping channels generate spin-independent harmonic currents due to the filled $p$ band. Observation of spin dependence therefore indicates a significant $d$–$d$ contribution, while its absence signals that charge-transfer processes dominate. The harmonic response thus directly identifies the microscopic hopping pathways of the material.

The harmonic spectrum also encodes the driving-field strength. As shown in Fig.~\ref{fig:hhg_compare}, the resonance condition $m\omega\approx U$ determines the interaction energy, while the plateau determines the field amplitude $\gamma=qEl/\omega$. Together, these features enable extraction of either material parameters or field intensity when the other is known.
\section{Conclusion and Outlook}
We have shown that higher-harmonic currents in Mott insulators depend sensitively on spin correlations, explaining recent experimental results that show this explicit dependence \cite{experiment}. Whereas, in charge-transfer insulators, the same behavior is further complicated by the dominating hopping channels and the initial band filling. The harmonic response therefore distinguishes between correlated electronic phases and directly reveals the dominant microscopic hopping pathways of the material.

We further demonstrated that harmonic spectra encode both interaction energy and driving-field strength. Higher harmonics thus provide a simultaneous probe of correlated materials and applied drives.

While related harmonic responses have already been observed experimentally \cite{anomalous, experiment}, our results suggest that future measurements in strongly correlated materials could directly extract spin correlations, hopping mechanisms, and interaction scales from nonlinear optical spectra, establishing higher-harmonic generation as a versatile sensing tool in the strongly interacting regime.
\section{Acknowledgment}
Funded by the Deutsche Forschungsgemeinschaft (DFG, German Research Foundation)
through the Collaborative Research Center SFB~1242
``Nonequilibrium dynamics of condensed matter in the time domain''
(Project-ID 278162697).
\bibliographystyle{apsrev4-2}
\bibliography{bib}
\end{document}